%% file: eprint.tex
\documentclass[12pt]{article}
\usepackage{graphicx}

\newcommand\pubnumber{}
\newcommand\pubdate{\today}

\def\uw{Wisconsin IceCube Particle Astrophysics Center, University of Wisconsin, Madison, WI 53706, U.S.A.}
\def\support{\footnote{http://icecube.wisc.edu}}

\def\Title#1{\begin{center} {\Large #1 } \end{center}}
\def\Author#1{\begin{center}{ \sc #1} \end{center}}
\def\Address#1{\begin{center}{ \it #1} \end{center}}

\newcommand\pubblock{\rightline{\begin{tabular}{l} \pubnumber\\
         \pubdate  \end{tabular}}}
\newenvironment{Abstract}{\begin{quotation}  }{\end{quotation}}
\newenvironment{Presented}{\begin{quotation} \begin{center} 
             PRESENTED AT\end{center}\bigskip 
      \begin{center}\begin{large}}{\end{large}\end{center} \end{quotation}}

\input econfmacros.tex

\begin{document}
\begin{titlepage}
\pubblock

\vfill
\Title{The Birth of Neutrino Astronomy}
\vfill
\Author{Naoko Kurahashi for the IceCube Collaboration\support}
\Address{\uw}
\vfill
\begin{Abstract}
First evidence of high-energy astrophysical neutrino observation with the IceCube detector from May 2010 to May 2012 is presented. 
Selecting for high-energy neutrino events with vertices well contained in the detector volume, the analysis has sensitivity starting at approximately 50 TeV, and the highest energy events are a few PeV.
A significant excess in flux is observed above expected backgrounds from atmospheric muons and neutrinos.
The sample of 28 events includes the highest energy neutrinos ever observed, and has properties consistent in flavor, arrival direction, and energy with generic expectations for neutrinos of extraterrestrial origin.
Using the new found astrophysical neutrinos, spatial searches of clusters are performed to look for exact sources, marking the birth of neutirno astronomy.
\end{Abstract}
\vfill
\begin{Presented}
Symposium on Cosmology and Particle Astrophysics\\
Honolulu, Hawai'i,  November 12--15, 2013
\end{Presented}
\vfill
\end{titlepage}
\def\thefootnote{\fnsymbol{footnote}}
\setcounter{footnote}{0}

\section{Introduction}

Observation of high-energy neutrinos are thought to provide insight into the origins and acceleration mechanisms of high-energy cosmic rays.
Cosmic-ray protons and nuclei produce neutrinos in interactions with gas in their source environment and interstellar dust through decay of charged pions and kaons.
These neutrinos have energies proportional to the cosmic rays that produced them and point back to their sources since they are neither affected by magnetic fields nor absorbed by matter opaque to radiation.
Neutrinos can also provide insight into the creation of gamma-rays at astronomical sources.
Gamma-rays can be produced by mesons from cosmic ray interactions in a similar process to neutrinos. 
This is known as the hadronic process, but they can also be produced by leptonic processes such as inverse compton scattering of electrons in high-energy source environments. 
Neutrino observation of gamma-ray sources will confirm their hadronic origin. 
Large-volume Cherenkov detectors like IceCube \cite{daqpaper} can detect these neutrinos through production of secondary leptons and hadronic showers when they interact with the detector material.
This analysis has an event selection with an energy threshold at about 50 TeV and is sensitive to energies up to 10 PeV and beyond, depending on the flux being measured. By selecting for events well contained in the detector volume, it is sensitive to all neutrino flavors from all directions.
Hints of astrophysical sources are seeked by characterizing the astrophysical flux measured and looking for spatial clustering among these events.

The analysis uses a data-taking period that started in May 2010 using 79 strings and continuing with the completed detector (86 strings) from May 2011 to May 2012 for a total livetime of 662 days.

\section{Event Selection}

Backgrounds for cosmic neutrino searches arise entirely from interactions of cosmic rays in the Earth's atmosphere.
These produce secondary muons that penetrate into underground neutrino detectors from above as well as atmospheric neutrinos that reach the detector from all directions due to the low neutrino cross-section which allows them to penetrate the Earth from the opposite hemisphere.
Neutrino candidates were selected by finding events that originated within the detector interior with sufficiently high energy such that an entering muon track would have been reliably identified if present.
This event selection rejects 99.999\% of the muon background while retaining approximately 98\% of all neutrino events interacting within the fiducial volume at energies above a few hundred TeV.
This selection is largely independent of neutrino flavor, event topology, or arrival direction.
It also removes 70\% of atmospheric neutrinos in the Southern Hemisphere, where atmospheric neutrinos are usually accompanied into the detector by muons produced in the same parent air shower.
For details, see \cite{HESEpaper} and \cite{icrc}.

Neutrino interactions in IceCube have two primary topologies: showers and muon tracks.
Secondary muon tracks are created in $\nu_\mu$ charged-current interactions and have a typical range that is on the order of kilometers, larger than the dimensions of the detector.
Showers are created by the secondary leptons produced in $\nu_e$ and $\nu_\tau$ charged-current interactions and in the neutral current interactions of neutrinos of all flavors.
At the relevant energies, showers have a length of roughly 10 meters in ice and are, to a good approximation, point sources of light.
Using the timing patterns of photon arrival times in individual PMTs allows for reconstruction of shower and track directions and deposited energies.
The typical median angular resolution for showers is $10^\circ$-$15^\circ$, whereas it is much better for tracks due to their extension (around $1^\circ$ or better, depending on their energy and length).

\section{Observed Data}
Twenty eight events are observed on an expected background of $10.6^{+5.0}_{-3.6}$ events from atmospheric muons and neutrinos. 
Seven events have clearly identifiable muon tracks, while the remaining 21 were shower-like.
Although there is some uncertainty in the expected atmospheric background rates, in particular from high-mass mesons with shorter lifetimes such as charm mesons, the energy spectrum, zenith distribution, and shower to muon track ratio of the observed events strongly disagree with the possibility that our events are entirely atmospheric in origin.

\begin{figure}[Htb]
\centering
\includegraphics[width=0.8\linewidth]{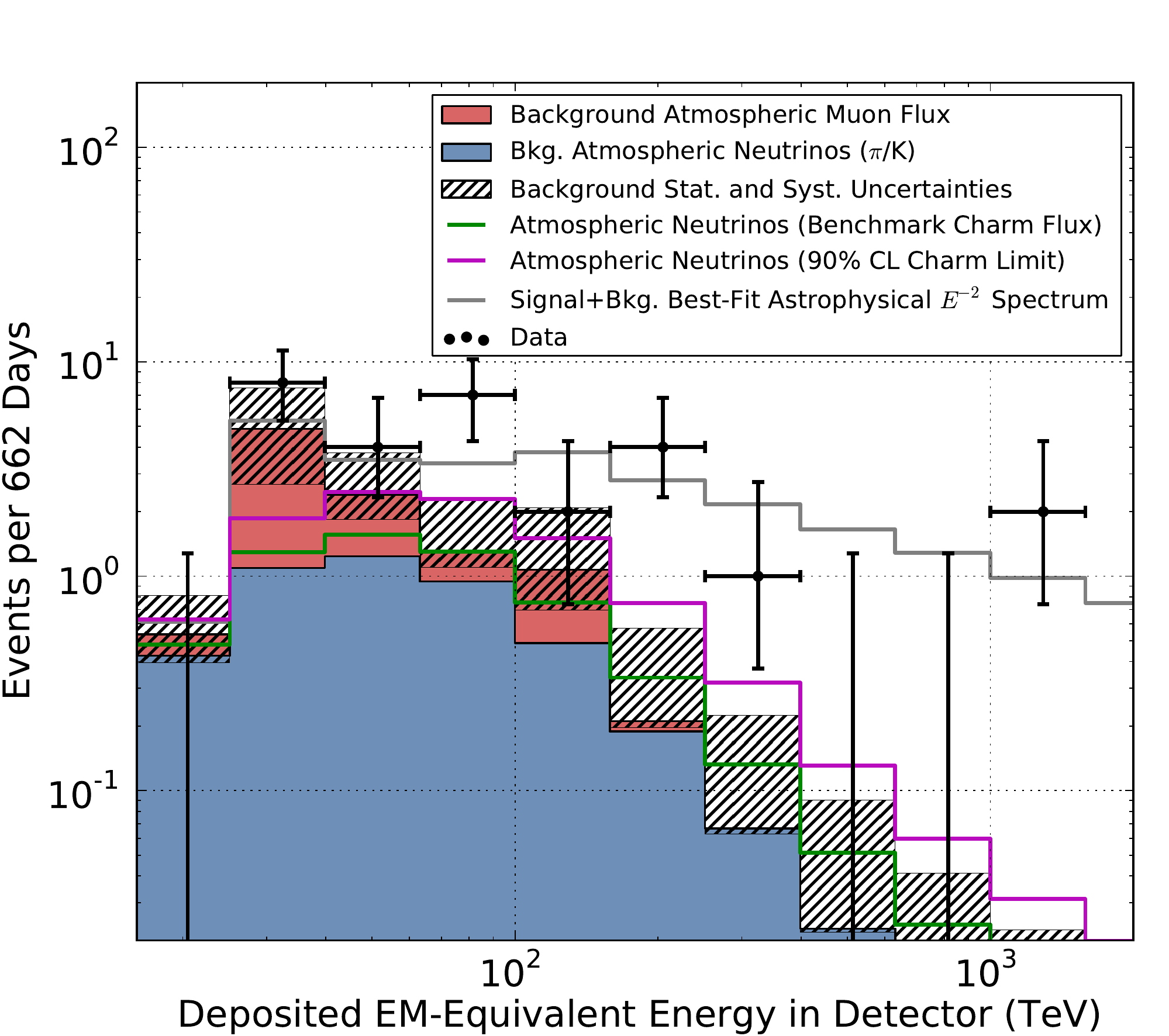}
\caption{
Distribution of the deposited energies of the observed events compared to model predictions.
Energies plotted are reconstructed in-detector visible energies, which are lower limits on the neutrino energy.
The expected rate of atmospheric neutrinos is shown in blue, with atmospheric muons in red.
The green line shows a benchmark atmospheric neutrino flux from charm mesons~\cite{charmmodel}, and the magenta line the experimental 90\% bound~\cite{charmlimit}.
Combined statistical and systematic uncertainties on the sum of backgrounds are indicated with a hatched area.
The gray line shows the best-fit $E^{-2}$ astrophysical spectrum with a per-flavor normalization (1:1:1) of $E^2 \Phi_{\nu}(E) = 1.2 \cdot 10^{-8}\, \mathrm{GeV}\, \mathrm{cm}^{-2}\, \mathrm{s}^{-1}\, \mathrm{sr}^{-1}$.
}
\label{fig:energyspectrum}
\end{figure}

The observed events are much higher in energy with a harder spectrum than expected from the atmospheric background, as seen in Fig.~\ref{fig:energyspectrum}.
Even the atmospheric neutrinos from charm mesons do not explain the excess without violating measured upper limits. The atmospheric muon background only contributes at lower energies and is unable to explain the high-energy excess.
By comparison, a neutrino flux produced in astrophysical sources is predicted have a harder spectrum like our data. Further discussions and observations can be found in \cite{HESEpaper}.

Three further characteristics about the astrophysical component can be inferred from the data. 
First is the compatibility to an equal neutrino flavor flux model. Because $6.0\pm 3.4$ track events are already expected from the muon background alone, the data is rich in shower-like events. An observed astrophysical flux will also be heavily biased toward showers because neutrino oscillations over astronomical baselines tend to equalize neutrino flavors \cite{2009PhRvD..80k3006C,2008JHEP...02..005P}.
Secondly, an unbroken $E^{-2}$ flux would have produced 3 to 6 more events in the 2-10~PeV range. Therefore, a slightly softer spectrum (best fit is $E^{-2.2}$) or a possible cut-off (best fit for a hard cut-off is $1.6^{+1.5}_{-0.4}$ PeV) explains the data better, but due to low statistics, a definitive statement cannot be made.
Thirdly, the astrophysical flux is consistent with an isotropic distribution. 
\begin{figure}[htb]
\centering
\includegraphics[width=0.48\linewidth]{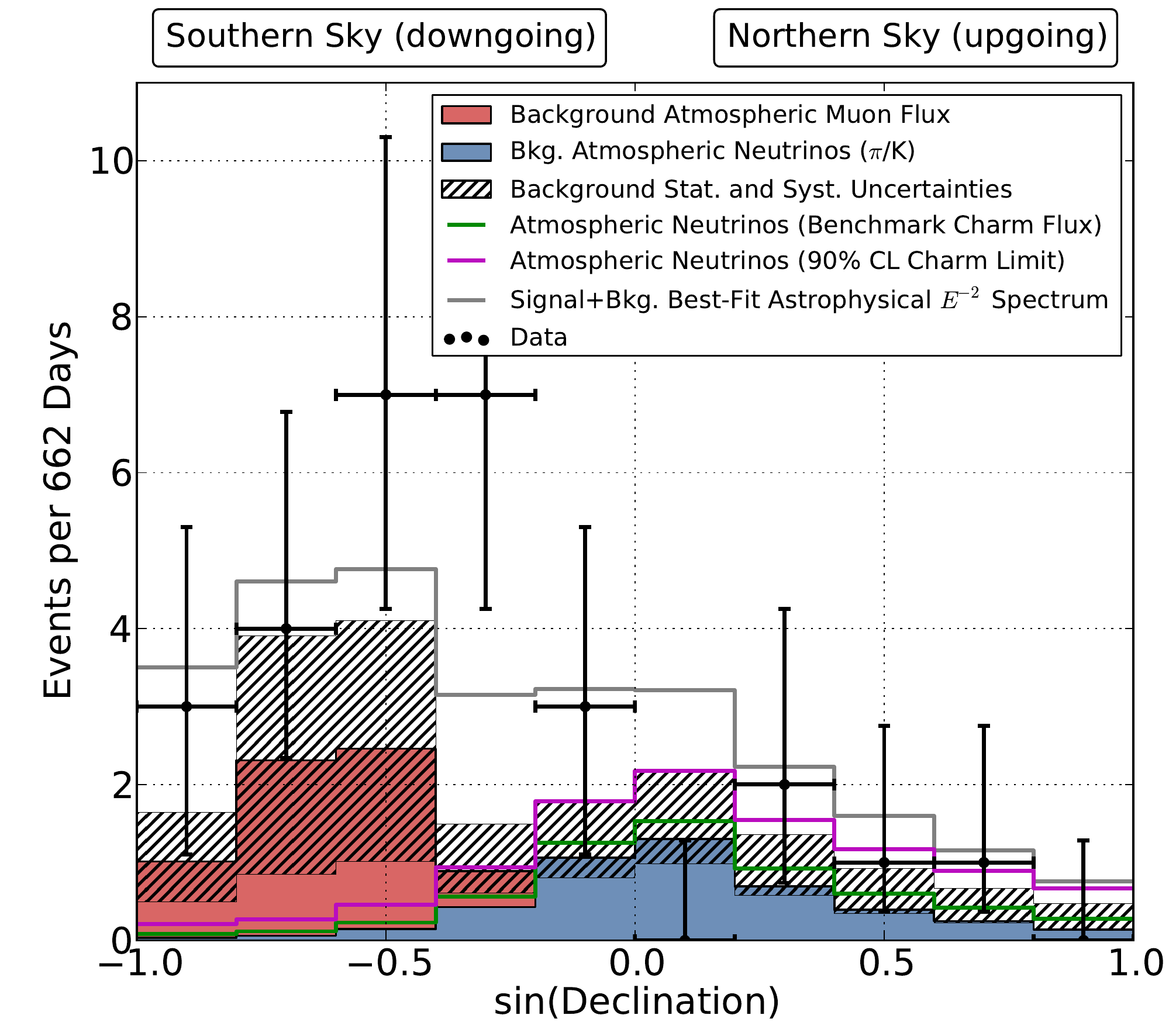}
\includegraphics[width=0.48\linewidth]{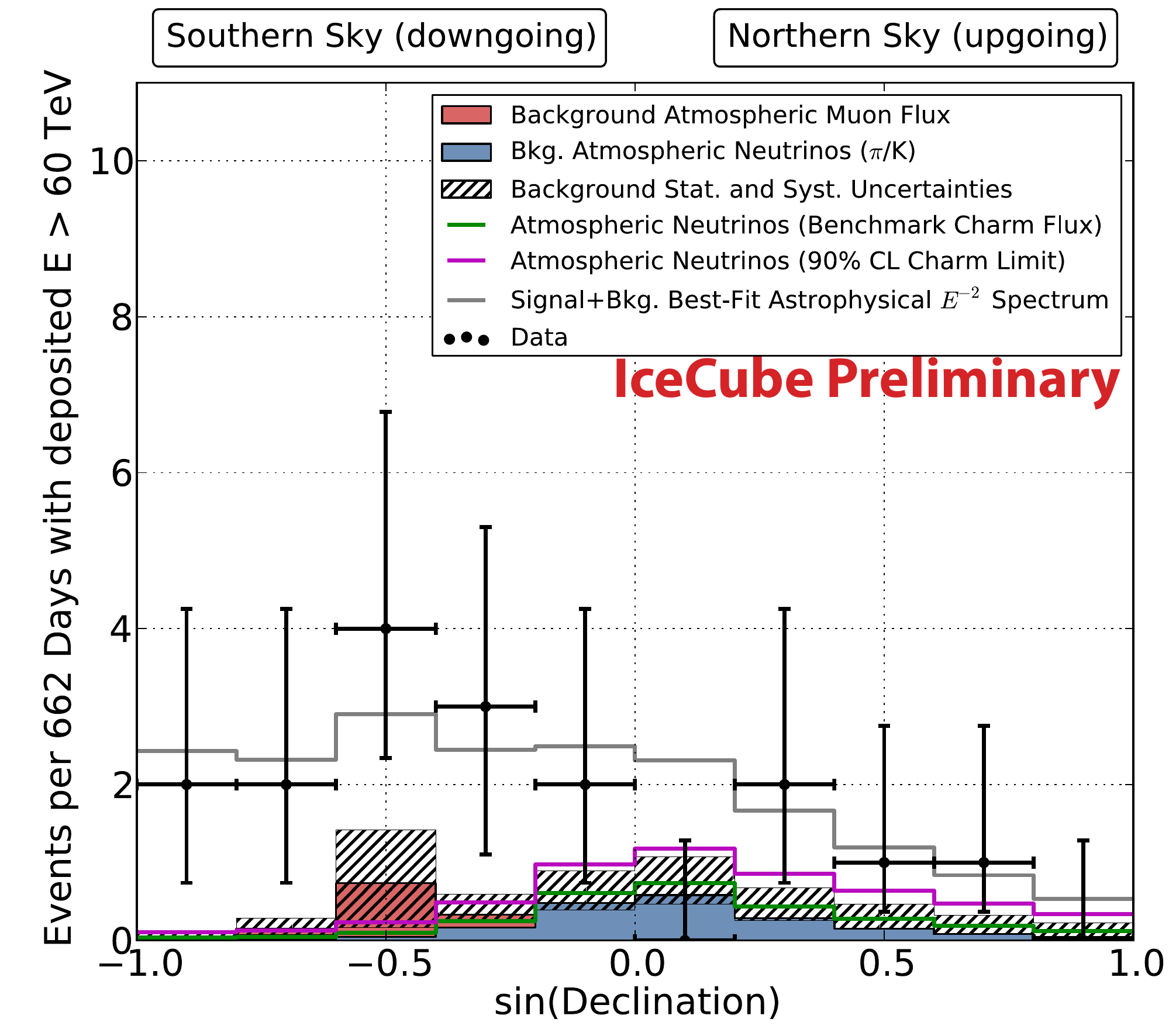}
\caption{
Distribution of the zenith angle of all observed events (left), and of those above 60 TeV in deposited energy (right) compared to model predictions.
Lines are the same as Fig.~\ref{fig:energyspectrum}. }
\label{fig:zen}
\end{figure}
Fig.~\ref{fig:zen} shows the zenith distribution of the observed events.
Atmospheric muons appear in the Southern Hemisphere as they cannot penetrate the Earth. 
Atmospheric neutrinos are surpressed in the vertically down-going region due to the accompanying shower muon triggering the veto. In the vertically up-going region, the absorption of neutrinos expected at these energies can be seen.
For the same reason, most events (approximately $60\%$, depending on the energy spectrum) from even an isotropic high-energy extraterrestrial population would be expected to appear in the Southern Hemisphere.

\section{Individual Source Search}
\begin{figure}[htb]
\centering
\includegraphics[width=0.9\linewidth]{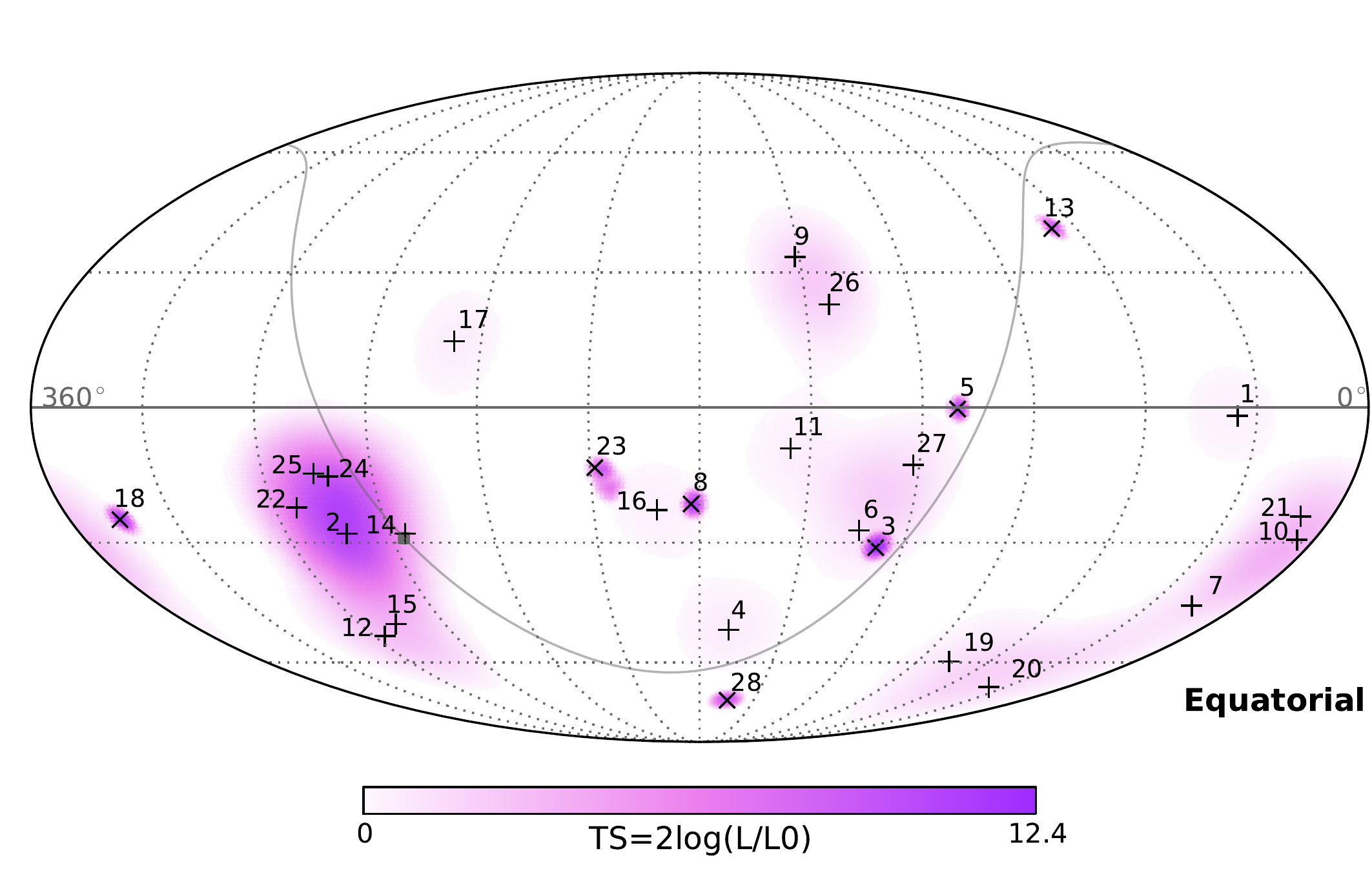}
\caption{Lines are the same as
Skymap in equatorial coordinates of the Test Statistic value (TS) from the maximum likelihood point-source analysis.
The galactic plane is shown as a gray line with the galactic center denoted as a filled gray square.
Best-fit locations of individual events (listed in~\cite{HESEpaper}) are indicated with vertical crosses ($+$) for showers and angled crosses ($\times$) for muon tracks.
}
\label{fig:ps}
\end{figure}

In order to test for spatial clustering of the events, a significance against the hypothesis that all events in this sample are uniformly distributed in right ascension was calculated~\cite{HESEpaper}.
Fig.~\ref{fig:ps} shows the result.
This test (performed once on the full sample and again on the subset of shower-like events) did not yield a significant result.
Several other tests (among them a galactic plane correlation study and multiple time clustering tests) did not yield significant results, either.

IceCube traditionally performs point source analyses using clean, throughgoing muons, and has limits from these analyses. The latest~\cite{jake} has a discovery potential at the level of approximately $E^2 \Phi_{\nu}(E) \sim 6 \cdot 10^{-9}\, \mathrm{GeV}\, \mathrm{cm}^{-2}\, \mathrm{s}^{-1}\ $ for the Northern Hemisphere, meaning this analysis would have discovered point sources above such flux. No such sources were seen. 
Comparing this flux to the observed isotropic astrophysical flux, it can be speculated that at least of order 10 sources must be contributing from the Northern Hemisphere. This speculation becomes invalid if the sources are extended in space or have significantly softer spectra.

One candidate sources to the astrophysical flux is starburst galaxies. IceCube also has limits on bright, close-by starburst galaxies using throughgoing muons~\cite{ic79ps}. For an unbroken $E^{-2}$ flux, the upper limit is 
$E^2 \Phi_{\nu}(E) \sim 9 \cdot 10^{-9}\, \mathrm{GeV}\, \mathrm{cm}^{-2}\, \mathrm{s}^{-1}\ $. Therefore, close-by starburst galaxies can only be responsible for up to $\sim 10\%$ of the astrophysical flux. However, it remains possible that most contributions to the flux are not the nearby starburst galaxies selected by the throughgoing muon analysis. The limit also assumes an unbroken $E^{-2}$ spectrum, which is not compatible with the observed flux.

\section{Conclusion}
An analysis of two years of IceCube data from 2010 to 2012 has observed a flux incompatible with expectations from terrestrial processes.
It contains a mixture of neutrino flavors compatible with a flux proportional to $E^{-2}$, a spectrum expected for neutrinos associated with primary cosmic ray acceleration.
The sample is thus consistent with generic expectations for a neutrino population with origins outside the solar system.
We did not observe significant spatial clustering of the events, although this study is currently limited by low statistics and poor angular resolution for the majority of the observed events.
Future observations with IceCube will provide improved measurements of the energy spectrum and origins of this flux, providing insight into the underlying processes responsible for these events.

\end{document}

%% file: econfmacros.tex
\def\beq{\begin{equation}}
\def\eeq#1{\label{#1}\end{equation}}
\def\eeqn{\end{equation}}

\def\beqa{\begin{eqnarray}}
\def\eeqa#1{\label{#1}\end{eqnarray}}
\def\eeqan{\end{eqnarray}}

\let\bar=\overbar

\def\Dslash{\not{\hbox{\kern-4pt $D$}}}
\def\dslash{\not{\hbox{\kern-2pt $\del$}}}

\def\msb{{\bar{\ssstyle M \kern -1pt S}}}

